\documentclass[conference]{IEEEtran}
\IEEEoverridecommandlockouts
\usepackage{cite}
\usepackage{amsmath,amssymb,amsfonts}
\usepackage{algorithmic}
\usepackage{graphicx}
\usepackage{textcomp}
\usepackage{xcolor}
\usepackage{pifont}
\usepackage{hyperref}
\usepackage{cleveref}

\def\BibTeX{{\rm B\kern-.05em{\sc i\kern-.025em b}\kern-.08em
    T\kern-.1667em\lower.7ex\hbox{E}\kern-.125emX}}
\begin{document}

\title{Boundary Vibration Control of Strain Gradient Timoshenko Micro-cantilevers Using Piezoelectric Actuators\\
}

\author{Amin Mehrvarz*, Hasan Salarieh*, Aria Alasty, and Ramin Vatankhah** \\
*Department of Mechanical Engineering, Sharif University of Technology \\
**Department of Mechanical Engineering, Shiraz University}

\maketitle

\begin{abstract}
In this paper, the problem of boundary control of vibration in a clamped-free strain gradient Timoshenko micro-cantilever is studied. For getting systems closer to reality, the force/moment exertion conditions should be modeled. To this end, a piezoelectric layer is laminated on one side of the beam and the controlling actuation is applied through the piezoelectric voltage. The beam and piezoelectric layer are coupled and modeled at the same time and the dynamic equations and boundary conditions of the system are achieved using the Hamilton principle. To achieve the purpose of eliminating vibration of the system, the control law is obtained from a Lyapunov function using LaSalle's invariant set theorem. The control law has a form of feedback from the spatial derivatives of boundary states of the beam. The finite element method using the strain gradient Timoshenko beam element has been used and then the simulation is performed to illustrate the impact of the proposed controller on the micro-beam.
\end{abstract}


\section{Introduction}
Many mechanical systems are modeled by partial differential equations with boundary conditions which are known as continues systems. To investigate such systems, accurate modeling is needed. This modeling is combined with a lot of simplification and as the level of simplification being lower, the model will being closer to reality\textcolor{blue}{\cite{wang2010micro, ramezani2012micro, gu2015analytical}}.

One of the important continuous systems in mechanical engineering is the micro-beam. Micro-beams have many applications in transportation\cite{mehrvarz2018modeling}, MEMS and NEMS\textcolor{blue}{\cite{mahmoodi2007non}},atomic force microscopy (AFM)\textcolor{blue}{\cite{indermuhle1997atomic}}, micro switches\textcolor{blue}{\cite{suzuki2003silicon, coutu2004comparison, vummidi2009dynamic, mohammadi2011mechanical}}, mass sensors, micro-accelerometers, micro-mirrors \textcolor{blue}{\cite{tuantranont2000smart, yee2000pzt}}, grating light valves (GLV)\textcolor{blue}{\cite{perry2004tomorrow, florin2010grating, rudra2010sige}} and cell contraction assays\textcolor{blue}{\cite{marzban2016effect}}. 

According to the position of continues systems control, various controllers have been designed for different purposes. For example in micro switches, the purpose of control is  position control of the end of beam\textcolor{blue}{\cite{mccarthy2002dynamic}}, in atomic force microscopy the goal is controlling the vibration of  the  beam in their resonant mode\textcolor{blue}{\cite{jalili2004review}} and in the static atomic force microscopy the goal is controlling the  shape and position of the beam\textcolor{blue}{\cite{krstic2008control}}.

Controller design methods are different. A group of these methods convert the partial differential equations (PDE) to several ordinary differential equations (ODE) and then for these equations, the controller is designed\textcolor{blue}{\cite{zhang2006adaptive, shirazi2013tip}}. In these methods, it is clear that the main system has been changed, and the infinite dimensions of the system are reduced to some finite dimensions of freedom, so the obtained controller is suitable when only a few specified modes of the system dynamics are excited which may not be guaranteed in real-world applications. Another group of controllers is boundary control methods that design the controller directly for the infinite dimensional main system and do not change it\textcolor{blue}{\cite{vatankhah2014asymptotic,vatankhah2013boundary, mehrvarz2018vibration}}. These controllers have many functions such as marine riser\textcolor{blue}{\cite{he2011adaptive, nguyen2013boundary, ardekany2016vibrations}} and robotic\textcolor{blue}{\cite{paranjape2013pde}}.

After designing the controller, it is required that control actuation being applied to the system. Various methods exist for this purpose. Electrostatic and piezoelectric actuators are the most common ones. One of the applications of electrostatic actuators is in grating light valves (GLV)\textcolor{blue}{\cite{maluf2002introduction}}. Besides, the piezoelectric actuators are usually utilized for atomic force microscopy, which nowadays is considered as the most effective tool in surface topography\textcolor{blue}{\cite{gaahlin1998novel, miyahara1999evaluation}} and also they are used for energy harvesting\textcolor{blue}{\cite{darabi2017analysis}}.

In this paper boundary control of a clamped-free strain gradient Timoshenko micro-cantilever with considering the effects of piezoelectric actuator is studied. In this state, it is assumed that a piezoelectric layer is ideally attached to one side of the beam. In the second section, the dynamic equations of the system are derived. A linear control law based on the theory of boundary control is proposed to suppress the system vibration, in section three. In the fourth section, the finite element method (FEM) is utilized for modeling the system. Simulation results before and after applying the control law are presented in the fifth section. Finally, the conclusion is given in the last section.

\section{Dynamics Model}
\begin{figure}
\centering
\includegraphics[width=80mm,scale=0.5]{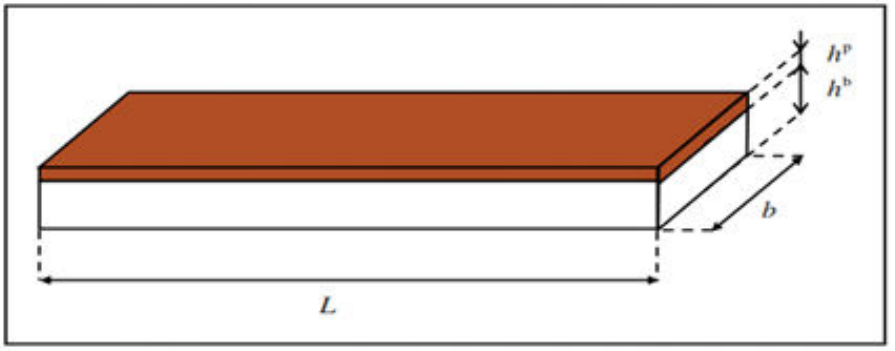}
\caption{A schematic view of the beam with piezoelectric actuator and some geometric parameters\textcolor{blue}{\cite{shirazi2013tip}}.}
\label{fig1}
\end{figure}

The investigated beam is a clamped-free strain gradient Timoshenko micro-cantilever that a layer of piezoelectric is ideally attached on it and shown in \textcolor{blue}{Fig.\ref{fig1}}. In this figure $h^b$ is the beam thickness, $h^p$ is the piezoelectric thickness, $L$ is the length of the beam and $b$ is the width of the beam.

The Hamilton principle is used to obtain partial differential equations with the boundary condition of the system that is shown in \eqref{eq1}.
\begin{align}
    \underset{{{t}_{0}}}{\overset{t}{\mathop \int }}\,\left( \delta T+\delta W_{e}^{*}-\delta {{W}_{m}}+\delta {{W}_{nc}} \right)dt=0
    \label{eq1}
\end{align}
where  $T$ is the kinetic energy, $W_{e}^{*}$  is the potential co-energy,   ${W}_{m}$ is the magnetic potential energy and ${W}_{nc}$  is the work of non-conservative forces\textcolor{blue}{\cite{preumont2006dynamics}}.

Also for the piezoelectric and beam, $W_{e}^{*}$  is obtained from equation \eqref{eq2}\textcolor{blue}{\cite{wang2010micro, preumont2006dynamics}}.
\begin{align}
\begin{split}
 & dW_{e}^{*}={{\left( \frac{dW_{e}^{*}}{dV} \right)}^{p}}d{{V}^{p}}+{{\left( \frac{dW_{e}^{*}}{dV} \right)}^{b}}d{{V}^{b}}= \left[\right. \left(\right. \frac{1}{2}{{E}^{T}}\varepsilon E\\ 
 & +{{S}^{T}}eE \left.\right)-\frac{1}{2}\left( {{\sigma }_{ij}}{{\varepsilon }_{ij}}+{{P}_{i}}{{\gamma }_{i}}+\tau _{jik}^{\left( 1 \right)}\eta _{ijk}^{\left( 1 \right)}+m_{ij}^{s}\chi _{ij}^{s} \right) \\ 
 & \times \left.\right]d{{V}^{p}}-\frac{1}{2}\left( {{\sigma }_{ij}}{{\varepsilon }_{ij}}+{{P}_{i}}{{\gamma }_{i}}+\tau _{jik}^{\left( 1 \right)}\eta _{ijk}^{\left( 1 \right)}+m_{ij}^{s}\chi _{ij}^{s} \right)d{{V}^{b}}  
\end{split}
\label{eq2}
\end{align}
where $E$ is the electric field vector that is shown in equation \eqref{eq3},   $\varepsilon$ is the permittivity matrix, $S$ is the strain vector in engineering representation that is shown in equation \eqref{eq4}, $e$  is the matrix of piezoelectric constants, ${\varepsilon }_{ij}$ is the strain tensor, ${\gamma }_{i}$  is the dilatation gradient tensor, $\eta _{ijk}^{\left( 1 \right)}$ is the deviatoric stretch gradient tensor and  $\chi _{ij}^{s}$ is the symmetric rotation gradient tensor. Also ${\sigma }_{ij}$ is the classical stress tensor, ${P}_{i}$، $\tau _{jik}$  and  $m_{ij}$ are higher-order stresses, $V$  is the total volume of material and superscript $^p$  and  $^b$ indicates that the regarded parameter is related to the piezo-layer or the beam. The relations given in \eqref{eq2} are written using the Einstein notation for summation.
\begin{align}
    & E={{\left[ \begin{matrix}
   0 & \begin{matrix}
   0 & {{E}_{3}}\left( t \right)  \\
\end{matrix}  \\
\end{matrix} \right]}^{T}}={{\left[ \begin{matrix}
   0 & \begin{matrix}
   0 & \frac{u\left( t \right)}{{{h}^{p}}}  \\
\end{matrix}  \\
\end{matrix} \right]}^{T}}
\label{eq3}
\\
& S={{\left[ \begin{matrix}
   -z{{\alpha }_{x}} & \begin{matrix}
   0 & \begin{matrix}
   0 & \begin{matrix}
   0 & \begin{matrix}
   2\beta  & 0  \\
\end{matrix}  \\
\end{matrix}  \\
\end{matrix}  \\
\end{matrix}  \\
\end{matrix} \right]}^{T}}
\label{eq4}
\end{align}

In the equation \eqref{eq3}, $u(t)$  is the piezoelectric voltage and in the equation \eqref{eq4}, $z$ is the distance from the neutral axis, $\alpha$  denotes the rotation of line elements along the centerline due to pure bending, and $\beta$  is obtained from the equation \eqref{eq5}.
\begin{align}
    {{v}_{x}}\left( x,t \right)=\beta \left( x,t \right)+\alpha \left( x,t \right)
    \label{eq5}
\end{align}
where $x$  and $t$  indicate the independent spatial (along the length of the beam) and time variables, respectively, $v$  represents the lateral deflection and subscripts $_x$  and $_t$  indicates derivative with respect to position and derivative with respect to time. The remaining equations are given from\textcolor{blue}{\cite{wang2010micro}}.

By replacing all the equation in \eqref{eq2}, we have:
\begin{align}
\begin{split}
 & W_{e}^{*}=\int\limits_{0}^{L}{[}\left( \frac{1}{2}{{\varepsilon }_{33}}\frac{{{u}^{2}}\left( t \right)}{{{h}^{p}}}-z{{e}_{13}}{{\alpha }_{x}}u\left( t \right) \right){{b}^{p}} \\ 
 & -\frac{1}{2}\left[\right. {{k}_{1}}^{p}{{\left( {{v}_{xxx}}-{{\beta }_{xx}} \right)}^{2}}+{{k}_{2}}^{p}{{\left( {{v}_{xx}}-{{\beta }_{x}} \right)}^{2}} \\ 
 & +{{k}_{3}}^{p}{{\left( 2{{v}_{xx}}-{{\beta }_{x}} \right)}^{2}}+{{k}_{4}}^{p}{{\left( {{v}_{xx}}-2{{\beta }_{x}} \right)}^{2}}+{{k}_{5}}^{p}{{\beta }^{2}} \left.\right]\\
 &-\frac{1}{2}\left[\right. {{k}_{1}}^{b}{{\left( {{v}_{xxx}}-{{\beta }_{xx}} \right)}^{2}}+{{k}_{2}}^{b}{{\left( {{v}_{xx}}-{{\beta }_{x}} \right)}^{2}} \\
 & +{{k}_{3}}^{b}{{\left( 2{{v}_{xx}}-{{\beta }_{x}} \right)}^{2}}+{{k}_{4}}^{b}{{\left( {{v}_{xx}}-2{{\beta }_{x}} \right)}^{2}}+{{k}_{5}}^{b}{{\beta }^{2}} \left.\right]]dx 
\end{split}
\label{eq6}
\end{align}
in the equation \eqref{eq6}, $k_i, i=1,2,...,5$  are defined as follows:
\begin{align}
  \left\{ \begin{matrix}
   \begin{matrix}
   {{k}_{1}}=\mu I\left( 2{{l}_{0}}^{2}+\frac{4}{5}{{l}_{1}}^{2} \right)  \\
   {{k}_{2}}=I\left( k+\frac{4}{3}\mu  \right)+2\mu A{{l}_{0}}^{2}  \\
\end{matrix}  \\
   {{k}_{3}}=\frac{1}{4}\mu A{{l}_{2}}^{2}  \\
   \begin{matrix}
   {{k}_{4}}=\frac{8}{15}\mu A{{l}_{1}}^{2}  \\
   {{k}_{5}}={{k}_{s}}\mu A  \\
\end{matrix}  \\
\end{matrix} \right.  
\label{eq7}
\end{align}
where $k$, $\mu$ and $k_s$ are the bulk module, shear module and the shear coefficient of the Timoshenko beam. $l_0$ , $l_1$ and $l_2$ demonstrate the additional independent material parameters

Defining the following parameters will simplify the governing equations.
\begin{align}
    \left\{ \begin{matrix}
   \begin{matrix}
   A={{\rho }^{p}}{{h}^{p}}{{b}^{p}}+{{\rho }^{p}}{{h}^{p}}{{b}^{p}}  \\
   \begin{matrix}
   B={{\rho }^{p}}{{I}^{p}}+{{\rho }^{b}}{{I}^{b}}  \\
   \begin{matrix}
   C={{k}_{1}}^{p}+{{k}_{1}}^{b}  \\
   D={{k}_{2}}^{p}+{{k}_{2}}^{b}  \\
\end{matrix}  \\
\end{matrix}  \\
\end{matrix} & \begin{matrix}
   E={{k}_{3}}^{p}+{{k}_{3}}^{b}  \\
   \begin{matrix}
   F={{k}_{4}}^{p}+{{k}_{4}}^{b}  \\
   \begin{matrix}
   G={{k}_{5}}^{p}+{{k}_{5}}^{b}  \\
   H=z{{e}_{13}}{{b}^{p}}  \\
\end{matrix}  \\
\end{matrix}  \\
\end{matrix}  \\
\end{matrix} \right.
\label{eq8}
\end{align}
where $\rho$ is the density of the beam or piezoelectric. 

The first variant of the potential co-energy  $W_{e}^{*}$, takes the following form:
\begin{align}
\begin{split}
 & \delta W_{e}^{*}=\underset{0}{\overset{L}{\mathop \int }}\,[-Hu\left( t \right)\delta {{\alpha }_{x}}-C{{\alpha }_{xx}}\delta {{\alpha }_{xx}}-D{{\alpha }_{x}}\delta {{\alpha }_{x}} -E\\ 
 & \times\left( {{v}_{xx}}+{{\alpha }_{x}} \right)\left( \delta {{v}_{xx}}+\delta {{\alpha }_{x}} \right)-F\left( 2{{\alpha }_{x}}-{{v}_{xx}} \right)\left(\right. 2\delta {{\alpha }_{x}} \\ 
 & -\delta {{v}_{xx}} \left.\right)-G\left( {{v}_{x}}-\alpha  \right)\left( \delta {{v}_{x}}-\delta \alpha  \right)]dx   
\end{split}
\label{eq9}
\end{align}

The kinetic energy of the system is obtained from the equation \eqref{eq10}.
\begin{align}
    T=\frac{1}{2}\underset{0}{\overset{L}{\mathop \int }}\,\left[ Av_{t}^{2}+B\alpha _{t}^{2} \right]dx
    \label{eq10}
\end{align}

The first variation of the kinetic energy $T$ is shown in the equation \eqref{eq11}.
\begin{align}
    \delta T=\underset{0}{\overset{L}{\mathop \int }}\,\left[ A{{v}_{t}}\delta {{v}_{t}}+B{{\alpha }_{t}}\delta {{\alpha }_{t}} \right]dx
    \label{eq11}
\end{align}

It is assumed that the external force is equal to zero.
\begin{align}
    {{W}_{m}}={{W}_{nc}}=0
    \label{eq12}
\end{align}

Replacing \eqref{eq9}, \eqref{eq11} and \eqref{eq12} in to \eqref{eq1}, we have:
\begin{align}
    \begin{split}
         & \underset{{{t}_{0}}}{\overset{t}{\mathop \int }}\,\underset{0}{\overset{L}{\mathop \int }}\,[A{{v}_{t}}\delta {{v}_{t}}+B{{\alpha }_{t}}\delta {{\alpha }_{t}}-Hu\left( t \right)\delta {{\alpha }_{x}}-C{{\alpha }_{xx}}\delta {{\alpha }_{xx}} \\ 
         & -D{{\alpha }_{x}}\delta {{\alpha }_{x}}-E\left( {{v}_{xx}}+{{\alpha }_{x}} \right)\left( \delta {{v}_{xx}}+\delta {{\alpha }_{x}} \right)-F\left(\right. 2{{\alpha }_{x}} \\ 
         & -{{v}_{xx}} \left.\right)\left( 2\delta {{\alpha }_{x}}-\delta {{v}_{xx}} \right)-G\left( {{v}_{x}}-\alpha  \right)\left( \delta {{v}_{x}}-\delta \alpha  \right)]dxdt \\
         &=0  
    \end{split}
    \label{eq13}
\end{align}

Using integration by parts on several terms of the equation (13), the following results are achieved:
\begin{align}
\begin{split}
        & \underset{0}{\overset{L}{\mathop \int }}\,\left[ \left. \left( A{{v}_{t}}\delta v+B{{\alpha }_{t}}\delta \alpha  \right) \right|_{{{t}_{0}}}^{t} \right]dx +\int\limits_{{{t}_{0}}}^{t}{\int\limits_{0}^{L}{[}}\left[\right. -A{{v}_{tt}}-E\\ 
         &\times\left( {{v}_{xxxx}}+{{\alpha }_{xxx}} \right)+F\left( 2{{\alpha }_{xxx}}-{{v}_{xxxx}} \right)+G\left(\right. {{v}_{xx}}- \\ 
         & {{\alpha }_{x}} \left.\right) \left.\right]\delta v+\left[\right. -B{{\alpha }_{tt}}-C{{\alpha }_{xxxx}}+D{{\alpha }_{xx}}+E\left(\right. {{v}_{xxx}}+ \\ 
         & {{\alpha }_{xx}} \left.\right)+2F\left( 2{{\alpha }_{xx}}-{{v}_{xxx}} \right)+G\left(\right. {{v}_{x}}-\alpha  \left.\right) \left.\right]\delta \alpha ]dxdt+  \\ 
         & \underset{{{t}_{0}}}{\overset{t}{\mathop \int }}\,[\left[\right. -Hu\left( t \right)+C{{\alpha }_{xxx}}-D{{\alpha }_{x}}-E\left( {{v}_{xx}}+{{\alpha }_{x}} \right)-2F \\ 
         &\times \left( 2{{\alpha }_{x}}-{{v}_{xx}} \right)+G\left( {{v}_{x}}-\alpha  \right) \left.\right]\delta \alpha+\left[ -C{{\alpha }_{xx}} \right]\delta {{\alpha }_{x}}+\left[\right. \\
         &  E\left( {{v}_{xxx}}+{{\alpha }_{xx}} \right)-F\left( 2{{\alpha }_{xx}}-{{v}_{xxx}} \right)-G\left( {{v}_{x}}-\alpha  \right) \left.\right]\delta v \\
         &+\left[ -E\left( {{v}_{xx}}+{{\alpha }_{x}} \right)+F\left( 2{{\alpha }_{x}}-{{v}_{xx}} \right) \right]\delta {{v}_{x}}]|_{0}^{L}dt=0 
\end{split}
\label{eq14}
\end{align}

Equation \eqref{eq14} is equal to zero. Therefore all of its terms should be equal to zero. So, the following equations are obtained:
\begin{align}
\begin{split}
    \left\{ \begin{matrix}
    \left. A{{v}_{t}}\delta v \right|_{{{t}_{0}}}^{t}=0  \\
    \left. B{{\alpha }_{t}}\delta \alpha  \right|_{{{t}_{0}}}^{t}=0  \\
    \end{matrix} \right.
\end{split}
\label{eq15}
\\
\begin{split}
\left\{ \begin{matrix}
  & A{{v}_{tt}}+\left( E+F \right){{v}_{xxxx}}+\left( E-2F \right){{\alpha }_{xxx}}-G\left(\right. {{v}_{xx}} \\ 
  &-{{\alpha }_{x}} \left.\right)=0 \\ 
 & B{{\alpha }_{tt}}+C{{\alpha }_{xxxx}}-\left( E-2F \right){{v}_{xxx}}-\left( D+E+4F \right) \\ 
 &\times {{\alpha }_{xx}}-G\left( {{v}_{x}}-\alpha  \right)=0 
 \end{matrix}\right.
\end{split}
\label{eq16}
\\
\begin{split}
 \left\{ \begin{matrix}
  & {{\left. ~\left( \left( E+F \right){{v}_{xxx}}+\left( E-2F \right){{\alpha }_{xx}}-G\left( {{v}_{x}}-\alpha  \right) \right) \right|}_{\left( L,t \right)}}\\
  &=0 \\ 
 & {{\left. ~\left( \left( E+F \right){{v}_{xx}}+\left( E-2F \right){{\alpha }_{x}} \right) \right|}_{\left( L,t \right)}}=0 \\ 
 & {{\left. ~~\left( C{{\alpha }_{xxx}}-\left( E-2F \right){{v}_{xx}}-\left( D+E+4F \right){{\alpha }_{x}} \right) \right|}_{\left( L,t \right)}} \\
 &=Hu\left( t \right) \\ 
 & {{\left. {{\alpha }_{xx}} \right|}_{\left( L,t \right)}}=0 \\ 
 & {{\left. v \right|}_{\left( 0,t \right)}}={{\left. {{v}_{x}} \right|}_{\left( 0,t \right)}}={{\left. \alpha  \right|}_{\left( 0,t \right)}}={{\left. {{\alpha }_{x}} \right|}_{\left( 0,t \right)}} \\ 
\end{matrix} \right.
\end{split}
\label{eq17}
\end{align}

Equation \eqref{eq16} is the partial differential equation of the system and equation \eqref{eq17} is the boundary condition of the system.
Potential co-energy of the system by using (8) will be obtained as follow:
\begin{align}
\begin{split}
 & U=\frac{1}{2}\underset{0}{\overset{L}{\mathop \int }}\,[{{\varepsilon }_{33}}\frac{{{b}^{p}}}{{{h}^{p}}}{{u}^{2}}\left( t \right)-H{{\alpha }_{x}}u\left( t \right)-C{{\left( {{v}_{xxx}}-{{\beta }_{xx}} \right)}^{2}} \\ 
 & -D{{\left( {{v}_{xx}}-{{\beta }_{x}} \right)}^{2}}-E{{\left( 2{{v}_{xx}}-{{\beta }_{x}} \right)}^{2}}-F{{\left( {{v}_{xx}}-2{{\beta }_{x}} \right)}^{2}} \\
 &-G{{\beta }^{2}}]dx
\end{split}
 \label{eq18}
\end{align}

In the next section, a boundary controller will be designed for the obtained model.

\section{Controller design}
Many flexible systems are modeled using a linear PDE and a set of BCs. To achieve the control purposes of flexible structures, most engineers rely on discretizing the governing PDE into a set of ordinary differential equations (ODEs)\textcolor{blue}{\cite{mehrvarz2018modeling, khodaei2018theoretical, ghaffarivardavagh2018horn}}. This is because of the abundance of control design techniques available for ODEs and mathematical complexities of boundary control of PDE models. For more clarification, in the field of vibration control of micro-cantilever beams, in references\textcolor{blue}{\cite{zhang2006adaptive,shirazi2013tip}}, the Galerkin method and finite element method were employed to change the governing PDE of the Euler–Bernoulli micro-beam to a set of ODEs, respectively. After that, a controller was designed for the resulting ODE model. 
Unfortunately, a stability result generated for a discretized ODE model under a proposed control cannot be generalized to the PDE model under the same control. That is, the neglected higher order modes could possibly destabilize the mechanical system under a discretized model-based control (i.e. spillover instability). Also, some devices and instruments such as strain gages are needed to feedback the vibration information at different points of the object, and an observer should be used to estimate the required vibration information based on the measured data. However, in many applications, using the measurement instruments at the interior points of the objects is impossible or at least very difficult\textcolor{blue}{\cite{najafi2010asymptotic}}. 

To eliminate the problems of both observation and control spillover, many investigators have proposed boundary control strategies for PDE models of elastic systems (i.e. the control involves only a few actuators placed at the boundary of media). The boundary controllers designed for the non-discretized PDE models are often simple compensators which ensure closed-loop stability for an infinite number of modes. The most significant advantage of the boundary control is that it can stabilize the motion of mechanical systems without using in-domain aligned actuators. This novelty has an important role in the field of industry and engineering applications.

\subsection{Boundary control}
If any undesired initial condition or noisy excitation is applied to the beam, the system may show unwanted vibrations. In this case, the boundary controller is designed to suppress the vibration and return the system to the equilibrium state. In our design, feedback of boundary states is utilized and the voltage of piezoelectric is tuned based on the feedback to stabilize the vibration.

Well-posedness of the closed-loop system for eliminating the system vibration has great importance. Semigroup technique and operator theory for designing the controller should be used. After that, benefitting from the Lyapunov stability method and the LaSalle's invariant set theorem, the asymptotic stability of the closed-loop system will be proved.
For boundary controller design and well-posedness analysis of the controlled system, the PDE model \eqref{eq16} should be written in the state-space, as shown in \eqref{eq19}.
\begin{align}
    {{X}_{t}}={{\left[ A \right]}_{4\times 4}}X
    \label{eq19}
\end{align}
where  $X$ is defined as:
\begin{align}
    X=\left[ \begin{matrix}
   v  \\
   \begin{matrix}
   {{v}_{t}}  \\
   \begin{matrix}
   \alpha   \\
   {{\alpha }_{t}}  \\
\end{matrix}  \\
\end{matrix}  \\
\end{matrix} \right]
\label{eq20}
\end{align}

Also, the matrix $A$ is defined as follow:
\begin{align}
    \left[ A \right]=\begin{matrix}
   \left[ \begin{matrix}
   \begin{matrix}
   0  \\
   \begin{matrix}
   {{a}_{v}}  \\
   \begin{matrix}
   0  \\
   {{b}_{v}}  \\
\end{matrix}  \\
\end{matrix}  \\
\end{matrix} & \begin{matrix}
   1  \\
   \begin{matrix}
   0  \\
   \begin{matrix}
   0  \\
   0  \\
\end{matrix}  \\
\end{matrix}  \\
\end{matrix}  \\
\end{matrix} \right. & \begin{matrix}
   \begin{matrix}
   0  \\
   \begin{matrix}
   {{a}_{\alpha }}  \\
   {{\begin{matrix}
   0  \\
   b  \\
\end{matrix}}_{\alpha }}  \\
\end{matrix}  \\
\end{matrix} & \left. \begin{matrix}
   0  \\
   \begin{matrix}
   0  \\
   \begin{matrix}
   1  \\
   0  \\
\end{matrix}  \\
\end{matrix}  \\
\end{matrix} \right]  \\
\end{matrix}  \\
\end{matrix}
\label{eq21}
\end{align}
where
\begin{align}
  \left\{ \begin{matrix}
  & {{a}_{v}}=-\frac{\left( E+F \right)}{A}\frac{{{\partial }^{4}}}{\partial {{x}^{4}}}+\frac{G}{A}\frac{{{\partial }^{2}}}{\partial {{x}^{2}}} \\ 
 & {{a}_{\alpha }}=-\frac{\left( E-2F \right)}{A}\frac{{{\partial }^{3}}}{\partial {{x}^{3}}}-\frac{G}{A}\frac{\partial }{\partial x} \\ 
 & {{b}_{v}}=\frac{\left( E-2F \right)}{B}\frac{{{\partial }^{3}}}{\partial {{x}^{3}}}+\frac{G}{B}\frac{\partial }{\partial x} \\ 
 & {{b}_{\alpha }}=-\frac{C}{B}\frac{{{\partial }^{4}}}{\partial {{x}^{4}}}+\frac{\left( D+E+4F \right)}{B}\frac{{{\partial }^{2}}}{\partial {{x}^{2}}}-\frac{G}{B} \\ 
\end{matrix} \right.
\label{eq22}
\end{align}

In the equation \eqref{eq19}, the matrix $A$ is the PDE operator. To achieve the controlling purpose (eliminating the system vibration), proper functional space should be chosen and the corresponding inner product should be defined by using the kinetic energy and potential co-energy without the terms of the electrical energy.

The proper functional space is denoted by $V$ which is defined on the proper functional space $(\Omega )$ and is shown in \eqref{eq23}.
\begin{align}
    V={{H}^{2}}\left( \text{ }\!\!\Omega\!\!\text{ } \right)\times {{L}^{2}}\left( \text{ }\!\!\Omega\!\!\text{ } \right)\times {{H}^{2}}\left( \text{ }\!\!\Omega\!\!\text{ } \right)\times {{L}^{2}}\left( \text{ }\!\!\Omega\!\!\text{ } \right)
    \label{eq23}
\end{align}

In \eqref{eq23}, ${L}^{p}(\Omega)$  is a Lebesgue space which is the space of measurable functions whose ${L}^{p}$  norm is bounded, equation \eqref{eq24}, and   ${H}^{k}(\Omega)$ is a Hilbert space that is defined in a Sobolev space   ${W}^{2k}(\Omega)$ in \eqref{eq25}.
\begin{align}
    {{\left[ \underset{\text{ }\!\!\Omega\!\!\text{ }}{\overset{~}{\mathop \int }}\,{{\left| f \right|}^{p}}d\mu  \right]}^{\frac{1}{p}}}<\infty 
    \label{eq24}
\end{align}
\begin{align}
\begin{split}
        &{{W}^{k2}}\left( \text{ }\!\!\Omega\!\!\text{ } \right)\equiv {{H}^{k}}\left( \text{ }\!\!\Omega\!\!\text{ } \right)= \\
        &\left\{ f:{{D}^{\alpha }}f\in {{L}^{2}}\left( \text{ }\!\!\Omega\!\!\text{ } \right),for~all~0\le \alpha \le k \right\}
\end{split}
\label{eq25}
\end{align}

In the equation \eqref{eq25}, ${D}^{\alpha}f$ is the $\alpha$th-order weak derivative of function\textcolor{blue}{\cite{reddy1986applied}}. 

The corresponding inner product introduced on the Hilbert space $V$ has the following form:
\begin{align}
\begin{split}
  & \left\langle Y,Z \right\rangle =\frac{1}{2}\underset{\text{ }\!\!\Omega\!\!\text{ }}{\overset{~}{\mathop \int }}\,[A{{a}_{2}}{{b}_{2}}+B{{a}_{4}}{{b}_{4}}+C{{a}_{3xx}}{{b}_{3xx}}+D{{a}_{3x}}{{b}_{3x}} \\ 
 & +E\left( {{a}_{1xx}}+{{a}_{3x}} \right)\left( {{b}_{1xx}}+{{b}_{3x}} \right)+F\left( 2{{a}_{3x}}-{{a}_{1xx}} \right)\left(\right. 2{{b}_{3x}} \\
 &-{{b}_{1xx}} \left.\right)+G\left( {{a}_{1x}}-{{a}_{3}} \right)\left( {{b}_{1x}}-{{b}_{3}} \right)]d\text{ }\!\!\Omega\!\!\text{ }  
\end{split}
\label{eq26}
\end{align}

In \eqref{eq26}, $Y=(a_1,a_2,a_3,a_4)$, $Z=(b_1,b_2,b_3,b_4)$  and $i=1,...,4$  are scalar-valued functions defined on $\Omega$  which are defined in \eqref{eq27}.
\begin{align}
    {{a}_{j}},{{b}_{j}}\in {{H}^{2}}\left( \text{ }\!\!\Omega\!\!\text{ } \right),~j=1,3;~{{a}_{j}},{{b}_{j}}\in {{L}^{2}}\left( \text{ }\!\!\Omega\!\!\text{ } \right),\text{ }\!\!~\!\!\text{ j}=2,4
    \label{eq27}
\end{align}

As mentioned, the inner product defined as the summation of kinetic energy \eqref{eq10} and potential co-energy \eqref{eq18} without the terms which correspond to electrical energy. The target of this investigation is to show that the system \eqref{eq16} with boundary conditions \eqref{eq17} under boundary feedbacks appeared in the equation \eqref{eq28} is well-posed and have an asymptotic decay rate.
\begin{align}
    u\left( t \right)={{k}_{u}}{{\alpha }_{t}}\left( L \right)
    \label{eq28}
\end{align}
In the equation \eqref{eq28},  $k_u$ is the controller gain and has a positive value.

The system equations \eqref{eq16} with boundary condition \eqref{eq17} in the state space is summarized as \eqref{eq29}.

\begin{align}
    \left\{ \begin{matrix}
  & {{X}_{t}}=\left[A\right]X \\ 
 & {{\Gamma }_{x=0}}: v={{v}_{x}}=\alpha ={{\alpha }_{x}}=0 \\ 
 & {{\Gamma }_{x=L}}:\\
 &\left\{\begin{matrix}
  & \left( E+F \right){{v}_{xxx}}\left( L \right)+\left( E-2F \right){{\alpha }_{xx}}\left( L \right) \\
  &-G\left( {{v}_{x}}\left( L \right)-\alpha \left( L \right) \right)=0 \\ 
 & \left( E+F \right){{v}_{xx}}\left( L \right)+\left( E-2F \right){{\alpha }_{x}}\left( L \right)=0 \\ 
 & C{{\alpha }_{xxx}}\left( L \right)-\left( E-2F \right){{v}_{xx}}\left( L \right)- \\
 &\left( D+E+4F \right){{\alpha }_{x}}\left( L \right)=Hu\left( t \right) \\ 
 & {{\alpha }_{xx}}\left( L \right)=0 \\ 
\end{matrix} \right. \\ 
\end{matrix} \right.
\label{eq29}
\end{align}

From operator $A$  and boundary conditions of the system in \eqref{eq29}, the domain of the operator $A$ is determined as \eqref{eq30}.
\begin{align}
    D\left( A \right)=H_{{{\Gamma }_{0}}}^{4}\left( \text{ }\!\!\Omega\!\!\text{ } \right)\times {{H}^{2}}\left( \text{ }\!\!\Omega\!\!\text{ } \right)\times H_{{{\Gamma }_{0}}}^{4}\left( \text{ }\!\!\Omega\!\!\text{ } \right)\times {{H}^{2}}\left( \text{ }\!\!\Omega\!\!\text{ } \right)
    \label{eq30}
\end{align}
where
\begin{align}
    H_{{{\Gamma }_{0}}}^{4}\left( \text{ }\!\!\Omega\!\!\text{ } \right)=\left\{ \left. f:f\in {{H}^{4}}\left( \text{ }\!\!\Omega\!\!\text{ } \right),\text{ }\!\!~\!\!\text{ f }\!\!~\!\!\text{ }{{|}_{{{\Gamma }_{0}}}}={{f}_{x}}~{{|}_{{{\Gamma }_{0}}}} \right\} \right.
    \label{eq31}
\end{align}

To illustrate the well-posedness of the controlled system expressed in \eqref{eq29}, first it should be proved that the operator $A$  is a dissipative operator.

$\boldsymbol{Theorem.1}$ The linear operator $A$  whose domain is defined in the equation \eqref{eq30} is dissipative.

$\boldsymbol{Proof.}$ From the definition of the inner product in the equation \eqref{eq26}, a Lyapunov function is defined as \eqref{eq32}.
\begin{align}
\begin{split}
& {{\left\langle X,X \right\rangle }_{V}}=\frac{1}{2}\underset{0}{\overset{L}{\mathop \int }}\,[A{{v}_{t}}^{2}+B{{\alpha }_{t}}^{2}+C{{\alpha }_{xx}}^{2}+D{{\alpha }_{x}}^{2} \\ 
 & +E{{\left( {{v}_{xx}}+{{\alpha }_{x}} \right)}^{2}}+F{{\left( 2{{\alpha }_{x}}-{{v}_{xx}} \right)}^{2}}+G{{\left( {{v}_{x}}-\alpha  \right)}^{2}} \\
 &]dx=E\left( t \right) 
\end{split}
\label{eq32} 
\end{align}

By taking the time derivatives of the Lyapunov function (32), we have:
\begin{align}
\begin{split}
 & \frac{d}{dt}{{\left\langle X,X \right\rangle }_{V}}=2{{\left\langle X,AX \right\rangle }_{V}}=\underset{0}{\overset{L}{\mathop \int }}\,[A{{v}_{t}}{{v}_{tt}}+B{{\alpha }_{t}}{{\alpha }_{tt}}+ \\ 
 & C{{\alpha }_{xx}}{{\alpha }_{xxt}}+D{{\alpha }_{x}}{{\alpha }_{xt}}+E\left( {{v}_{xx}}+{{\alpha }_{x}} \right)\left( {{v}_{xxt}}+{{\alpha }_{xt}} \right)+ \\
 &F\left(\right. 2{{\alpha }_{x}}-{{v}_{xx}} \left.\right)\left( 2{{\alpha }_{xt}}-{{v}_{xxt}} \right)+G\left( {{v}_{x}}-\alpha  \right)\left(\right. {{v}_{xt}}- \\
 &{{\alpha }_{t}} \left.\right)]dx
\end{split}
  \label{eq33}
\end{align}

By replacing  ${v}_{tt}$ and  ${\alpha }_{tt}$  form equation \eqref{eq16}, the following equation is obtained:

\begin{align}
\begin{split}
  & {{\left\langle X,AX \right\rangle }_{V}}=\frac{1}{2}\underset{0}{\overset{L}{\mathop \int }}\,[-{{v}_{t}}\left[\right. \left( E+F \right){{v}_{xxxx}}+\left( E-2F \right) \\ 
 & \times{{\alpha }_{xxx}}-G\left( {{v}_{xx}}-{{\alpha }_{x}} \right) \left.\right]-{{\alpha }_{t}}\left[\right. C{{\alpha }_{xxxx}}-\left( E-2F \right) \\ 
 &\times{{v}_{xxx}}-\left( D+E+4F \right){{\alpha }_{xx}}-G\left( {{v}_{x}}-\alpha  \right) \left.\right]+C{{\alpha }_{xx}} \\
 &{{\alpha }_{xxt}}+D{{\alpha }_{x}}{{\alpha }_{xt}} +E\left( {{v}_{xx}}+{{\alpha }_{x}} \right)\left( {{v}_{xxt}}+{{\alpha }_{xt}} \right)+F \left(\right. 2\\
 &\times{{\alpha }_{x}}-{{v}_{xx}} \left.\right)\left( 2{{\alpha }_{xt}}-{{v}_{xxt}} \right)+G\left( {{v}_{x}}-\alpha  \right)\left( {{v}_{xt}}-{{\alpha }_{t}} \right)\\
 &]dx
\end{split}
\label{eq34}
\end{align}

By rearranging \eqref{eq34}, we have
\begin{align}
\begin{split}
& {{\left\langle X,AX \right\rangle }_{V}}=\frac{1}{2}\int\limits_{0}^{L}{[}E\left( {{v}_{xx}}{{v}_{xxt}}-{{v}_{xxxx}}{{v}_{t}} \right)+F\left(\right. {{v}_{xx}} \\ 
 & \times{{v}_{xxt}}-{{v}_{xxxx}}{{v}_{t}} \left.\right)+G\left( {{v}_{xx}}{{v}_{t}}+{{v}_{x}}{{v}_{xt}} \right)+E\left(\right. {{v}_{xxx}} \\ 
 & \times{{\alpha }_{t}}-{{\alpha }_{xxx}}{{v}_{t}}+{{\alpha }_{xt}}{{v}_{xx}}+{{\alpha }_{x}}{{v}_{xxt}} \left.\right)+2F\left(\right. {{\alpha }_{xxx}}{{v}_{t}} \\ 
 & -{{v}_{xxx}}{{\alpha }_{t}}-{{\alpha }_{x}}{{v}_{xxt}}-{{\alpha }_{xt}}{{v}_{xx}} \left.\right)+G\left(\right. {{\alpha }_{t}}{{v}_{x}}-{{\alpha }_{x}}{{v}_{t}}\\
 &-{{\alpha }_{t}}{{v}_{x}}-\alpha {{v}_{xt}} \left.\right)+C\left(\right. {{\alpha }_{xx}}{{\alpha }_{xxt}}-{{\alpha }_{xxxx}}{{\alpha }_{t}} \left.\right)+D\left(\right. \\
 &{{\alpha }_{xx}}{{\alpha }_{t}}+ {{\alpha }_{x}}{{\alpha }_{xt}} \left.\right)+E\left(\right. {{\alpha }_{xx}}{{\alpha }_{t}}+{{\alpha }_{x}}{{\alpha }_{xt}} \left.\right)+4F\left(\right. {{\alpha }_{xx}} \\
 &\times{{\alpha }_{t}}+{{\alpha }_{x}}{{\alpha }_{xt}}\left.\right)]dx
\end{split}
 \label{eq35} 
\end{align}

Performing some integration by parts on \eqref{eq35}, the following results are achieved:
\begin{align}
\begin{split}
 & {{\left\langle X,AX \right\rangle }_{V}}=\frac{1}{2}[E\left( -{{v}_{xxx}}{{v}_{t}}+{{v}_{xx}}{{v}_{xt}} \right)+F\left(\right. -{{v}_{xxx}}{{v}_{t}}\\
 & +{{v}_{xx}}{{v}_{xt}} \left.\right)+G{{v}_{x}}{{v}_{t}}+E\left( {{\alpha }_{t}}{{v}_{xx}}+{{\alpha }_{x}}{{v}_{xt}}-{{\alpha }_{xx}}{{v}_{t}} \right)+2 \\
 & \times F\left( {{\alpha }_{xx}}{{v}_{t}}-{{\alpha }_{x}}{{v}_{xt}}-{{\alpha }_{t}}{{v}_{xx}} \right)-G\alpha {{v}_{t}}+C\left(\right. -{{\alpha }_{xxx}}{{\alpha }_{t}} \\
 &+{{\alpha }_{xx}}{{\alpha }_{xt}} \left.\right)+D{{\alpha }_{x}}{{\alpha }_{t}}+E{{\alpha }_{x}}{{\alpha }_{t}}+4F{{\alpha }_{x}}{{\alpha }_{t}}]\mathop{|}_{0}^{L}   
\end{split}
\label{eq36}
\end{align}

Factorizing the terms that have time derivatives results in 
\begin{align}
\begin{split}
 & {{\left\langle X,AX \right\rangle }_{V}}=\frac{1}{2}[-{{v}_{t}}\left(\right. E{{v}_{xxx}}+F{{v}_{xxx}}-G{{v}_{x}}+E{{\alpha }_{xx}}\\ 
 & -2F{{\alpha }_{xx}}+G\alpha  ) +{{v}_{xt}}\left( E{{v}_{xx}}+F{{v}_{xx}}+E{{\alpha }_{x}}-2F{{\alpha }_{x}} \right) \\ 
 & +{{\alpha }_{t}}\left( E{{v}_{xx}}-2F{{v}_{xx}}-C{{\alpha }_{xxx}}+D{{\alpha }_{x}}+E{{\alpha }_{x}}+4F{{\alpha }_{x}} \right)\\
 &+C{{\alpha }_{xt}}{{\alpha }_{xx}}]\mathop{|}_{0}^{L}  
\end{split}
\label{eq37}
\end{align}

Implementing boundary condition \eqref{eq17} into equation \eqref{eq37} yields,
\begin{align}
\begin{split}
  & \left\langle X,A{{X}_{V}} \right\rangle =\frac{1}{2}[-{{v}_{t}}\left(\right. \left( E+F \right){{v}_{xxx}}+\left( E-2F \right){{\alpha }_{xx}}\\ 
 & -G\left( {{v}_{x}}-\alpha  \right) \left.\right) +{{v}_{xt}}\left( \left( E+F \right){{v}_{xx}}+\left( E-2F \right){{\alpha }_{x}} \right) \\ 
 &-{{\alpha }_{t}}~\left( C{{\alpha }_{xxx}}-\left( E-2F \right){{v}_{xx}}-\left( D+E+4F \right){{\alpha }_{x}} \right) \\ &+C{{\alpha }_{xt}}{{\alpha }_{xx}}]\mathop{|}_{0}^{L}     
\end{split}
\label{eq38} 
\end{align}

By replacing \eqref{eq17} in \eqref{eq38} we have,
\begin{align}
{{\left\langle X,AX \right\rangle }_{V}}=-\frac{1}{2}H{{\alpha }_{t}}\left( L \right)u\left( t \right)
    \label{eq39}
\end{align}

By replacing control law \eqref{eq28} in \eqref{eq39} the following relation is obtained.
\begin{align}
{{\left\langle X,AX \right\rangle }_{V}}=-\frac{1}{2}H{{k}_{u}}{{\alpha }_{t}}^{2}\left( L \right)
    \label{eq40}
\end{align}

According to \eqref{eq40}, it is clear that for the closed-loop system we have,
\begin{align}
{{\left\langle X,AX \right\rangle }_{V}}\le 0
    \label{eq41}
\end{align}

Thus, from the definition of the dissipative operators\textcolor{blue}{\cite{robinson2001infinite}}, the proof will be complete.\ding{110}

In the following, the continuity of the operator $(\gamma I-A)^{-1}$ is checked to achieve the final purpose. 

$\boldsymbol{Theorem.2}$ The operator $(\gamma I-A)^{-1}$  exists and it is continuous for any $\gamma$.

$\boldsymbol{Proof.}$ It is assumed that we have:
\begin{align}
    \left( \gamma I-A \right)X={{X}_{0}}
    \label{eq42}
\end{align}

For demonstrating the existence of the operator $(\gamma I-A)^{-1}$ it is sufficient to show that only one solution exists for \eqref{eq42}. The result of theorem 1 (equation \eqref{eq40}) is used to obtain the following relation. 
\begin{align}
\begin{split}
    & {{\left\langle \left( \gamma I-A \right)X,X \right\rangle }_{V}}={{\left\langle \gamma X,X \right\rangle }_{V}}-{{\left\langle AX,X \right\rangle }_{V}}= \\ 
 & \gamma {{\left\langle X,X \right\rangle }_{V}}+\frac{1}{2}H{{k}_{u}}{{\alpha }_{t}}^{2}\left( L \right)\ge \gamma {{\left\langle X,X \right\rangle }_{V}}=\gamma \left\| X \right\|_{V}^{2} 
\end{split}
  \label{eq43}
\end{align}

The above result is shown that the bilinear form $q$ with the definition of $a(u,v)=\left\langle(\gamma I-A)u,v\right\rangle$  is coercive on the Hilbert space $V$. Now, using the Lax-Milgram theorem, one can easily prove equation 42  has a unique weak solution and so the operator   exists $(\gamma I-A)^{-1}$\textcolor{blue}{\cite{yosida1971functional}}.

In\textcolor{blue}{\cite{reddy1986applied}} it is shown that if the operator $(\gamma I-A)^{-1}$  is bounded, it will be continuous. So considering equation \eqref{eq43} that is obtained from the dissipativity of operator   one can conclude that:
\begin{align}
\begin{split}
        & \gamma \left\| X \right\|_{V}^{2}\le {{\left\langle \left( \gamma I-A \right)X,X \right\rangle }_{V}}={{\left\langle {{X}_{0}},X \right\rangle }_{V}} \\
        & \le {{\left\| {{X}_{0}} \right\|}_{V}}{{\left\| X \right\|}_{V}}\to ~{{\left\| {{X}_{0}} \right\|}_{V}}\ge \gamma {{\left\| X \right\|}_{V}}
\end{split}
    \label{eq44}
\end{align}

Since $\left\| {{X}_{0}} \right\|_V$ is bounded, $\left\|X\right\|_V$ is also bounded and as a result, because boundedness contains the continuity, the operator $(\gamma I-A)^{-1}$ is continues and the proof is complete.\ding{110}

In the following, according to the control purpose we show that equation \eqref{eq29} is well-posed.

$\boldsymbol{Theorem.3}$ Equation \eqref{eq29} with initial condition  $X(t=0) \in D(A)$ is well-posed.

$\boldsymbol{Proof.}$ According to the definition of the functional domain space, that is ${{H}^{4}}(\Omega )\times {{H}^{2}}(\Omega )\times {{H}^{4}}(\Omega )\times {{H}^{2}}(\Omega )\subset V$, it is clear that  $D(A)$ is dense in $V$. Also, it is clear that the range of $(\gamma I-A)$ is dense in $V$, it means:
\begin{align}
    \overline{\mathcal{R}\left( \gamma I-A \right)}=V
    \label{eq45}
\end{align}

According to theorem 2, $(\gamma I-A)$ has a continuous inverse $(\gamma I-A)^{-1}$ for any $\gamma>0$. Therefore, according to the definition of the resolving set of an operator\textcolor{blue}{\cite{robinson2001infinite}}, $\gamma$ is in the resolving set of the operator $A$.

As shown in theorem 1, it is demonstrated that the operator $A$  is a dissipative operator. Therefore, according to the Lumer-Phillips theorem\textcolor{blue}{\cite{pazy2012semigroups}}, equation \eqref{eq29} with control law \eqref{eq28} and initial condition $X(t=0) \in D(A)$ is well-posed.\ding{110} 

The asymptotic stability of the closed-loop system is achieved by using LaSalle’s invariant set theorem which is based on the Lyapunov method. For using this theorem, it should be shown that $(\gamma I-A)^{-1}$  is compact for any  $\gamma>0$\textcolor{blue}{\cite{luo2012stability}}.

$\boldsymbol{Theorem.4}$ Operator $(\gamma I-A)^{-1}$  is compact for any  $\gamma>0$.

$\boldsymbol{Proof.}$ It is shown that the operator $(\gamma I-A)^{-1}$  for any  $\gamma>0$ is bounded. This subject is shown in the proof of theorem 2. Also, it is obvious that:

\begin{align}
    {{\left( \gamma I-A \right)}^{-1}}V~\subset D\left( A \right)
    \label{eq46}
\end{align}

According to Rellich-Kondrachov compact embedding theorem\textcolor{blue}{\cite{robinson2001infinite}}, since the closure of $(\gamma I-A)^{-1}V$ is ${{H}^{4}}(\Omega )\times {{H}^{2}}(\Omega )\times {{H}^{4}}(\Omega )\times {{H}^{2}}(\Omega )$  and this space is compactly embedded in ${{H}^{2}}(\Omega )\times {{L}^{2}}(\Omega )\times {{H}^{2}}(\Omega )\times {{L}^{2}}(\Omega )$ \textcolor{blue}{\cite{robinson2001infinite}}, therefore the compactness of the above-mentioned resolving is obtained and the proof will be completed.\ding{110}

According to these theorems, by using LaSalle's invariant set theorem, the asymptotic stability of the closed loop system will be demonstrated.

$\boldsymbol{Theorem.5}$ The system of equation \eqref{eq29} with control feedback \eqref{eq28} will asymptotically tend toward zero.

$\boldsymbol{Proof.}$ According to selected Lyapunov function that contains some terms of kinetic energy and potential co-energy and according to the defined inner product, it was shown that $E(t)=\left\langle {X,X} \right\rangle_V \ge 0 $ is positive definite. Also, it was shown in Theorem 1 that the time derivative of Lyapunov function is equal to:
\begin{align}
    \dot{E}\left( t \right)=-{{k}_{1}}{{k}_{u}}{{\alpha }_{t}}^{2}\left( L \right)
    \label{eq47}
\end{align}

Equation \eqref{eq47} shows only the convergence of $\alpha_t(t)$  to zero, but one can use the LaSalle theorem to prove the asymptotic stability. So, Theorems 2-4 have been proved.  It is clear from the above equation that $\dot{E}(t)\le 0$  and $E(t)\ge 0$ has requirements of a Lyapunov function. Therefore, Because of compactness of the resolving $(\gamma I-A)^{-1}$  proved in Theorem 4, the LaSalle’s invariant set theorem\textcolor{blue}{\cite{luo2012stability}} gives the asymptotic decay rate of the controlled and the proof will be complete.

\section{Finite Element Method}
In this section for system modeling, the finite element method is provided. For modeling, strain gradient Timoshenko beam element is selected. It is assumed that the element has two nodes and each node has  $\left[\begin{matrix}v&v_x&\alpha&\alpha_x \end{matrix}\right]$ variables and   and   have the following polynomial forms:
\begin{align}
    \left\{ \begin{matrix}
   v={{c}_{1}}+{{c}_{2}}x+{{c}_{3}}{{x}^{2}}+{{c}_{4}}{{x}^{3}}  \\
   \alpha ={{c}_{5}}+{{c}_{6}}x+{{c}_{7}}{{x}^{2}}+{{c}_{8}}{{x}^{3}}  \\
\end{matrix} \right.
\label{eq48}
\end{align}

For using this element, at first the shape function should be obtained and then with using the shape function, kinetic and potential energy, mass, stiffness and force matrices are calculated. 

Equation \eqref{eq48} can be written in a matrix form as \eqref{eq49}.
\begin{align}
    \left[ \begin{matrix}
   v  \\
   \begin{matrix}
   {{v}_{x}}  \\
   \begin{matrix}
   \alpha   \\
   {{\alpha }_{x}}  \\
\end{matrix}  \\
\end{matrix}  \\
\end{matrix} \right]=\left[ \begin{matrix}
   \begin{matrix}
   1  \\
   \begin{matrix}
   0  \\
   \begin{matrix}
   0  \\
   0  \\
\end{matrix}  \\
\end{matrix}  \\
\end{matrix} & \begin{matrix}
   \begin{matrix}
   x  \\
   \begin{matrix}
   1  \\
   \begin{matrix}
   0  \\
   0  \\
\end{matrix}  \\
\end{matrix}  \\
\end{matrix} & \begin{matrix}
   \begin{matrix}
   {{x}^{2}}  \\
   \begin{matrix}
   2x  \\
   \begin{matrix}
   0  \\
   0  \\
\end{matrix}  \\
\end{matrix}  \\
\end{matrix} & \begin{matrix}
   \begin{matrix}
   {{x}^{3}}  \\
   \begin{matrix}
   3{{x}^{2}}  \\
   \begin{matrix}
   0  \\
   0  \\
\end{matrix}  \\
\end{matrix}  \\
\end{matrix} & \begin{matrix}
   \begin{matrix}
   0  \\
   \begin{matrix}
   0  \\
   \begin{matrix}
   1  \\
   0  \\
\end{matrix}  \\
\end{matrix}  \\
\end{matrix} & \begin{matrix}
   \begin{matrix}
   0  \\
   \begin{matrix}
   0  \\
   \begin{matrix}
   x  \\
   1  \\
\end{matrix}  \\
\end{matrix}  \\
\end{matrix} & \begin{matrix}
   \begin{matrix}
   0  \\
   \begin{matrix}
   0  \\
   \begin{matrix}
   {{x}^{2}}  \\
   2x  \\
\end{matrix}  \\
\end{matrix}  \\
\end{matrix} & \left. \begin{matrix}
   0  \\
   \begin{matrix}
   0  \\
   \begin{matrix}
   {{x}^{3}}  \\
   3{{x}^{2}}  \\
\end{matrix}  \\
\end{matrix}  \\
\end{matrix} \right]  \\
\end{matrix}  \\
\end{matrix}  \\
\end{matrix}  \\
\end{matrix}  \\
\end{matrix}  \\
\end{matrix}  \\
\end{matrix} \right.\left[ \begin{matrix}
   {{c}_{1}}  \\
   \begin{matrix}
   {{c}_{2}}  \\
   \begin{matrix}
   {{c}_{3}}  \\
   \begin{matrix}
   {{c}_{4}}  \\
   \begin{matrix}
   {{c}_{5}}  \\
   \begin{matrix}
   {{c}_{6}}  \\
   \begin{matrix}
   {{c}_{7}}  \\
   {{c}_{8}}  \\
\end{matrix}  \\
\end{matrix}  \\
\end{matrix}  \\
\end{matrix}  \\
\end{matrix}  \\
\end{matrix}  \\
\end{matrix} \right]=gC
\label{eq49}
\end{align}
where in \eqref{eq49}, $c_i,i=1,...,8$ are constant and $h$ matrix for two nodes $x=0,L_e$  has been calculated that $L_e$ is the length of the beam element.
\begin{align}
    h=\left[ \begin{matrix}
   \begin{matrix}
   1  \\
   \begin{matrix}
   0  \\
   \begin{matrix}
   0  \\
   \begin{matrix}
   0  \\
   \begin{matrix}
   1  \\
   \begin{matrix}
   0  \\
   \begin{matrix}
   0  \\
   0  \\
\end{matrix}  \\
\end{matrix}  \\
\end{matrix}  \\
\end{matrix}  \\
\end{matrix}  \\
\end{matrix}  \\
\end{matrix} & \begin{matrix}
   \begin{matrix}
   0  \\
   \begin{matrix}
   1  \\
   \begin{matrix}
   0  \\
   \begin{matrix}
   0  \\
   \begin{matrix}
   {{L}_{e}}  \\
   \begin{matrix}
   1  \\
   \begin{matrix}
   0  \\
   0  \\
\end{matrix}  \\
\end{matrix}  \\
\end{matrix}  \\
\end{matrix}  \\
\end{matrix}  \\
\end{matrix}  \\
\end{matrix} & \begin{matrix}
   \begin{matrix}
   0  \\
   \begin{matrix}
   0  \\
   \begin{matrix}
   0  \\
   \begin{matrix}
   0  \\
   \begin{matrix}
   {{L}_{e}}^{2}  \\
   \begin{matrix}
   2{{L}_{e}}  \\
   \begin{matrix}
   0  \\
   0  \\
\end{matrix}  \\
\end{matrix}  \\
\end{matrix}  \\
\end{matrix}  \\
\end{matrix}  \\
\end{matrix}  \\
\end{matrix} & \begin{matrix}
   \begin{matrix}
   0  \\
   \begin{matrix}
   0  \\
   \begin{matrix}
   0  \\
   \begin{matrix}
   0  \\
   \begin{matrix}
   {{L}_{e}}^{3}  \\
   \begin{matrix}
   3{{L}_{e}}^{2}  \\
   \begin{matrix}
   0  \\
   0  \\
\end{matrix}  \\
\end{matrix}  \\
\end{matrix}  \\
\end{matrix}  \\
\end{matrix}  \\
\end{matrix}  \\
\end{matrix} & \begin{matrix}
   \begin{matrix}
   0  \\
   \begin{matrix}
   0  \\
   \begin{matrix}
   1  \\
   \begin{matrix}
   0  \\
   \begin{matrix}
   0  \\
   \begin{matrix}
   0  \\
   \begin{matrix}
   1  \\
   0  \\
\end{matrix}  \\
\end{matrix}  \\
\end{matrix}  \\
\end{matrix}  \\
\end{matrix}  \\
\end{matrix}  \\
\end{matrix} & \begin{matrix}
   0  \\
   \begin{matrix}
   0  \\
   \begin{matrix}
   0  \\
   \begin{matrix}
   1  \\
   \begin{matrix}
   0  \\
   \begin{matrix}
   0  \\
   \begin{matrix}
   {{L}_{e}}  \\
   1  \\
\end{matrix}  \\
\end{matrix}  \\
\end{matrix}  \\
\end{matrix}  \\
\end{matrix}  \\
\end{matrix}  \\
\end{matrix} & \begin{matrix}
   \begin{matrix}
   0  \\
   \begin{matrix}
   0  \\
   \begin{matrix}
   0  \\
   \begin{matrix}
   0  \\
   \begin{matrix}
   0  \\
   \begin{matrix}
   0  \\
   \begin{matrix}
   {{L}_{e}}^{2}  \\
   2{{L}_{e}}  \\
\end{matrix}  \\
\end{matrix}  \\
\end{matrix}  \\
\end{matrix}  \\
\end{matrix}  \\
\end{matrix}  \\
\end{matrix} & \begin{matrix}
   0  \\
   \begin{matrix}
   0  \\
   \begin{matrix}
   0  \\
   \begin{matrix}
   0  \\
   \begin{matrix}
   0  \\
   \begin{matrix}
   0  \\
   \begin{matrix}
   {{L}_{e}}^{3}  \\
   3{{L}_{e}}^{2}  \\
\end{matrix}  \\
\end{matrix}  \\
\end{matrix}  \\
\end{matrix}  \\
\end{matrix}  \\
\end{matrix}  \\
\end{matrix}  \\
\end{matrix}  \\
\end{matrix}  \\
\end{matrix}  \\
\end{matrix}  \\
\end{matrix}  \\
\end{matrix} \right]
\label{eq50} \\
\left[ \begin{matrix}
   {{c}_{1}}  \\
   \begin{matrix}
   {{c}_{2}}  \\
   \begin{matrix}
   {{c}_{3}}  \\
   \begin{matrix}
   {{c}_{4}}  \\
   \begin{matrix}
   {{c}_{5}}  \\
   \begin{matrix}
   {{c}_{6}}  \\
   \begin{matrix}
   {{c}_{7}}  \\
   {{c}_{8}}  \\
\end{matrix}  \\
\end{matrix}  \\
\end{matrix}  \\
\end{matrix}  \\
\end{matrix}  \\
\end{matrix}  \\
\end{matrix} \right]={{h}^{-1}}\left[ \begin{matrix}
   \begin{matrix}
   {{v}_{1}}  \\
   \begin{matrix}
   {{v}_{1x}}  \\
   \begin{matrix}
   {{\alpha }_{1}}  \\
   {{\alpha }_{1x}}  \\
\end{matrix}  \\
\end{matrix}  \\
\end{matrix}  \\
   \begin{matrix}
   {{v}_{2}}  \\
   \begin{matrix}
   {{v}_{2x}}  \\
   \begin{matrix}
   {{\alpha }_{2}}  \\
   {{\alpha }_{2x}}  \\
\end{matrix}  \\
\end{matrix}  \\
\end{matrix}  \\
\end{matrix} \right]={{h}^{-1}}q
\label{eq51}
\end{align}

Finally:
\begin{align}
    \left[ \begin{matrix}
   v  \\
   \begin{matrix}
   {{v}_{x}}  \\
   \begin{matrix}
   \alpha   \\
   {{\alpha }_{x}}  \\
\end{matrix}  \\
\end{matrix}  \\
\end{matrix} \right]=g{{h}^{-1}}q=Nq
\label{eq52}
\end{align}
where $N$ in \eqref{eq52} is the shape function of a Timoshenko beam element that is defined as:
\begin{align}
    \begin{split}
    &N=g{{h}^{-1}} \\
    &=\left[ \begin{matrix}
   \begin{matrix}
   {{H}_{1}}  \\
   \begin{matrix}
   {{H}_{1x}}  \\
   \begin{matrix}
   0  \\
   0  \\
\end{matrix}  \\
\end{matrix}  \\
\end{matrix} & \begin{matrix}
   \begin{matrix}
   {{H}_{2}}  \\
   \begin{matrix}
   {{H}_{2x}}  \\
   \begin{matrix}
   0  \\
   0  \\
\end{matrix}  \\
\end{matrix}  \\
\end{matrix} & \begin{matrix}
   \begin{matrix}
   0  \\
   \begin{matrix}
   0  \\
   \begin{matrix}
   {{H}_{1}}  \\
   {{H}_{1x}}  \\
\end{matrix}  \\
\end{matrix}  \\
\end{matrix} & \begin{matrix}
   \begin{matrix}
   0  \\
   \begin{matrix}
   0  \\
   \begin{matrix}
   {{H}_{2}}  \\
   {{H}_{2x}}  \\
\end{matrix}  \\
\end{matrix}  \\
\end{matrix} & \begin{matrix}
   \begin{matrix}
   {{H}_{3}}  \\
   \begin{matrix}
   {{H}_{3x}}  \\
   \begin{matrix}
   0  \\
   0  \\
\end{matrix}  \\
\end{matrix}  \\
\end{matrix} & \begin{matrix}
   \begin{matrix}
   {{H}_{4}}  \\
   \begin{matrix}
   {{H}_{4x}}  \\
   \begin{matrix}
   0  \\
   0  \\
\end{matrix}  \\
\end{matrix}  \\
\end{matrix} & \begin{matrix}
   \begin{matrix}
   0  \\
   \begin{matrix}
   0  \\
   \begin{matrix}
   {{H}_{3}}  \\
   {{H}_{3x}}  \\
\end{matrix}  \\
\end{matrix}  \\
\end{matrix} & \begin{matrix}
   0  \\
   \begin{matrix}
   0  \\
   \begin{matrix}
   {{H}_{4}}  \\
   {{H}_{4x}}  \\
\end{matrix}  \\
\end{matrix}  \\
\end{matrix}  \\
\end{matrix}  \\
\end{matrix}  \\
\end{matrix}  \\
\end{matrix}  \\
\end{matrix}  \\
\end{matrix}  \\
\end{matrix} \right]
    \end{split}
    \label{eq53}
\end{align}

In \eqref{eq53}, $H_i,i=1,...,4$ are obtained as follows:
\begin{align}
    \left\{ \begin{matrix}
  & {{H}_{1}}=\frac{2{{x}^{3}}}{{{L}_{e}}^{3}}-\frac{3{{x}^{2}}}{{{L}_{e}}^{2}}+~1 \\ 
 & {{H}_{2}}=x-\frac{2{{x}^{2}}}{{{L}_{e}}}+\frac{{{x}^{3}}}{{{L}_{e}}^{2}} \\ 
 & {{H}_{3}}=\frac{3{{x}^{2}}}{{{L}_{e}}^{2}}-\frac{2{{x}^{3}}}{{{L}_{e}}^{3}} \\ 
 & {{H}_{4}}=\frac{{{x}^{3}}}{{{L}_{e}}^{2}}-\frac{2{{x}^{2}}}{{{L}_{e}}} \\ 
\end{matrix} \right.
\label{eq54}
\end{align}

Now by using kinetic energy and potential co-energy \eqref{eq10} and \eqref{eq18} and variational method\textcolor{blue}{\cite{huebner2008finite}}, mass, stiffness and force matrices will be obtained as \eqref{eq55}, \eqref{eq56} and \eqref{eq57}.

\begin{align}
&{{M}_{e}}=\underset{0}{\overset{{{L}_{e}}}{\mathop \int }}\,({{D}_{1}}^{T}A{{D}_{1}}+{{D}_{2}}^{T}B{{D}_{2}})dx 
\label{eq55} \\
\begin{split}
    &{{K}_{e}}=\underset{0}{\overset{{{L}_{e}}}{\mathop \int }}\,({{B}_{1}}^{T}C{{B}_{1}}+{{B}_{2}}^{T}D{{B}_{2}} \\
    &+{{B}_{3}}^{T}E{{B}_{3}}+{{B}_{4}}^{T}F{{B}_{4}}+{{B}_{5}}^{T}G{{B}_{5}})dx
\end{split}
\label{eq56}\\
&{{F}_{e}}=\underset{0}{\overset{{{L}_{e}}}{\mathop \int }}\,-\frac{1}{2}{{k}_{1}}UB_{2}^{T}dx
\label{eq57}
\end{align}

In the above equations  $D_i,B_j,i=1,2$ and $j=1,...,4$ are defined as:
\begin{align}
    \left\{ \begin{matrix}
  & {{D}_{1}}=\left[ \begin{matrix}
   \begin{matrix}
   1 & \begin{matrix}
   0 & 0  \\
\end{matrix}  \\
\end{matrix} & 0  \\
\end{matrix} \right]N \\ 
 & {{D}_{2}}=\left[ \begin{matrix}
   \begin{matrix}
   0 & 0  \\
\end{matrix} & \begin{matrix}
   1 & 0  \\
\end{matrix}  \\
\end{matrix} \right]N \\ 
 & {{B}_{1}}=\left[ \begin{matrix}
   \begin{matrix}
   0 & \begin{matrix}
   0 & 0  \\
\end{matrix}  \\
\end{matrix} & \frac{\partial }{\partial x}  \\
\end{matrix} \right]N \\ 
 & {{B}_{2}}=\left[ \begin{matrix}
   \begin{matrix}
   0 & \begin{matrix}
   0 & 0  \\
\end{matrix}  \\
\end{matrix} & 1  \\
\end{matrix} \right]N \\ 
 & {{B}_{3}}=\left[ \begin{matrix}
   \begin{matrix}
   0 & \begin{matrix}
   \frac{\partial }{\partial x} & 0  \\
\end{matrix}  \\
\end{matrix} & 1  \\
\end{matrix} \right]N \\ 
 & {{B}_{4}}=\left[ \begin{matrix}
   \begin{matrix}
   0 & \begin{matrix}
   -\frac{\partial }{\partial x} & 0  \\
\end{matrix}  \\
\end{matrix} & 2  \\
\end{matrix} \right]N \\ 
 & {{B}_{5}}=\left[ \begin{matrix}
   \begin{matrix}
   0 & 1  \\
\end{matrix} & \begin{matrix}
   -1 & 0  \\
\end{matrix}  \\
\end{matrix} \right]N \\ 
\end{matrix} \right.
\label{eq58}
\end{align}

For making a model of the real system by assuming ten nodes in the system, matrices $M$, $K$ and $F$ will be obtained by assembling the matrices given in equation \eqref{eq55}, \eqref{eq56} and \eqref{eq57}. Time evolution of the system will be obtained by numerical integration of the following equation.
\begin{align}
    \left[ M \right]\left\{ {\ddot{q}} \right\}+\left[ K \right]\left\{ q \right\}=\left\{ F \right\}
    \label{eq64}
\end{align}
\begin{figure}
\centering
\includegraphics[width=80mm,scale=0.5]{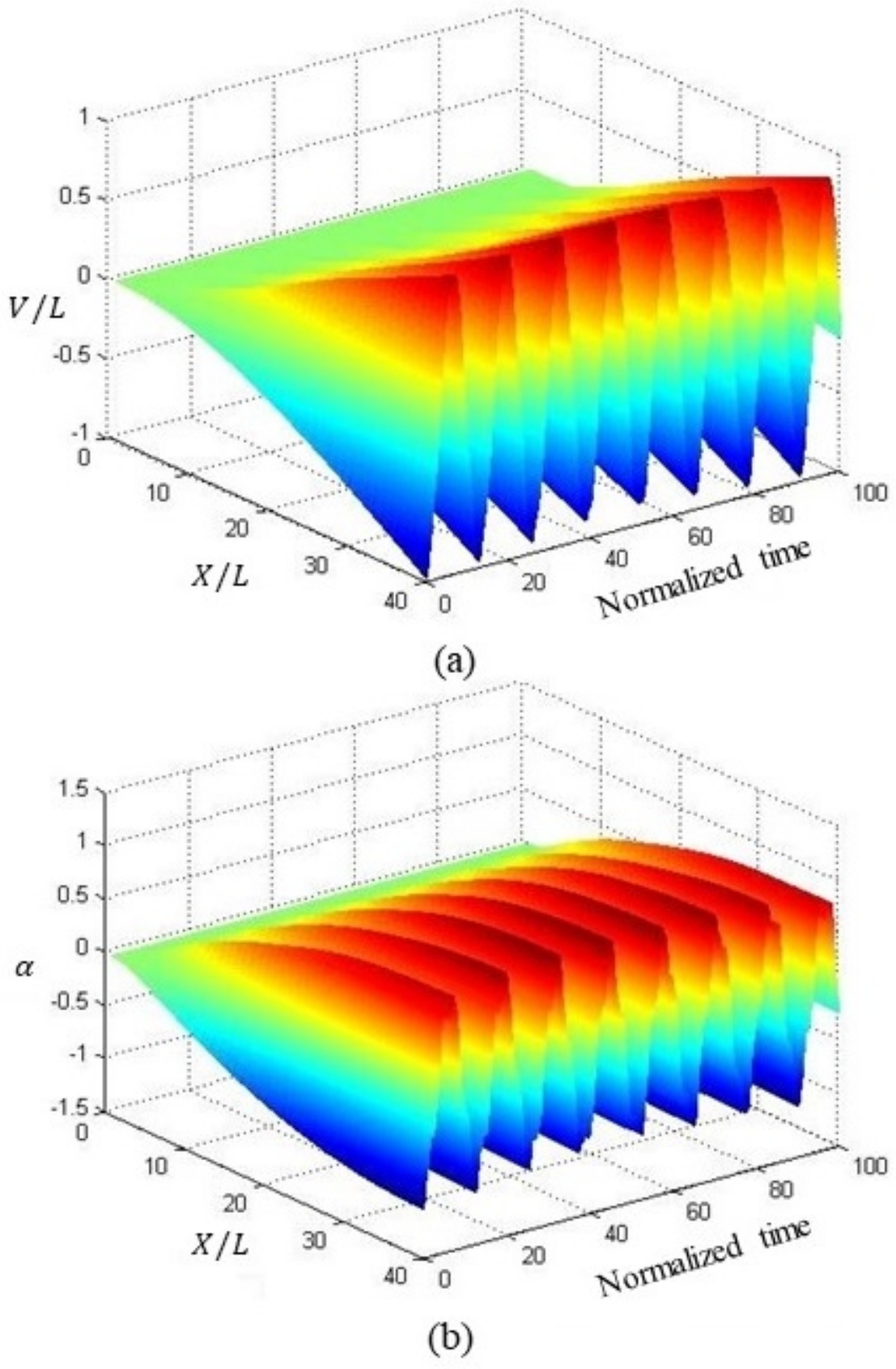}
\caption{Response of the micro beam before control voltage exertion: (a) lateral deflection $v(x,t)$, (b) rotation of line elements along the centerline  $\alpha(x,t)$.}
\label{fig2}
\end{figure}

\begin{figure}
\centering
\includegraphics[width=80mm,scale=0.5]{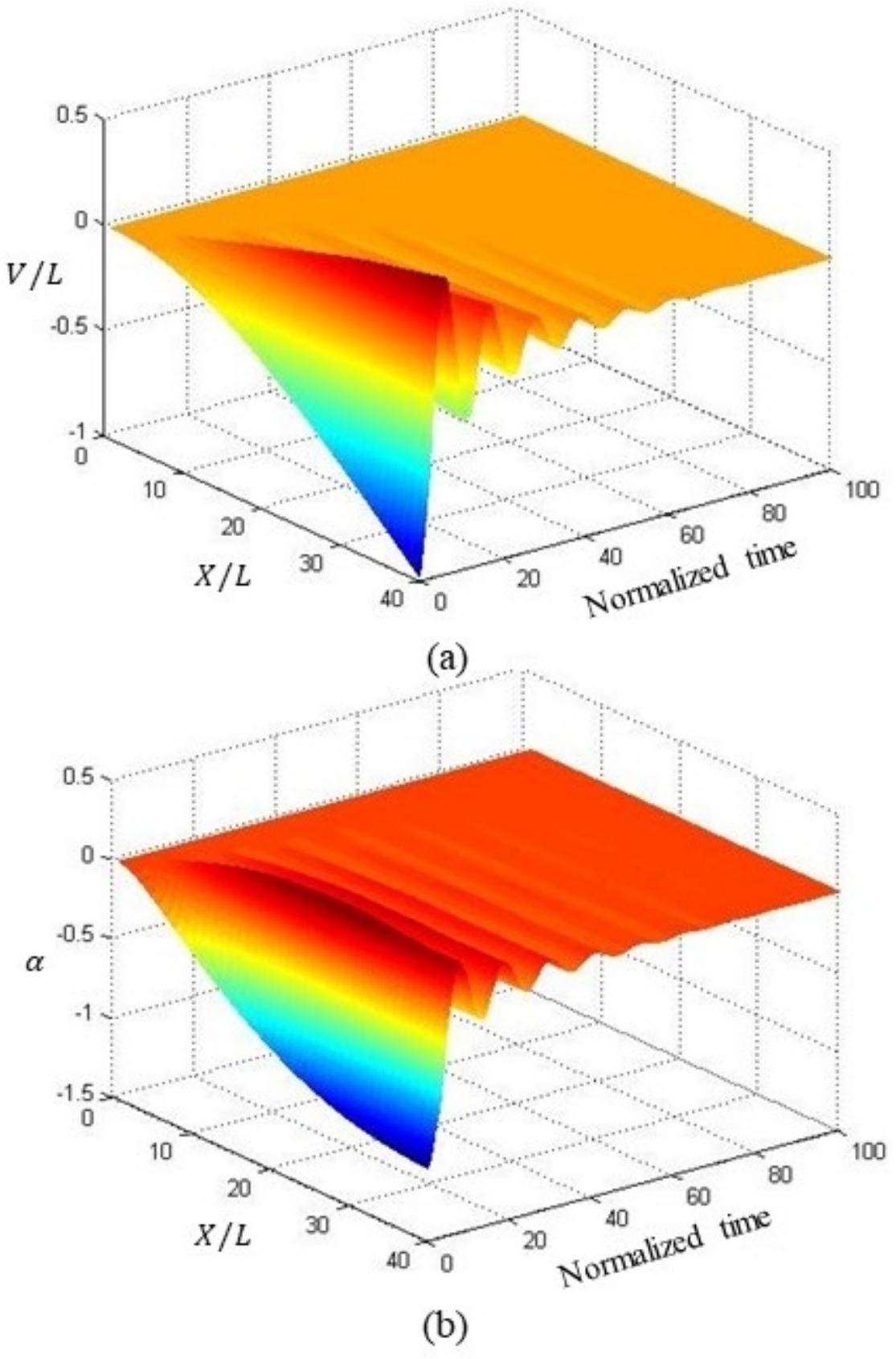}
\caption{Response of the micro beam after control voltage exertion: (a) lateral deflection $v(x,t)$, (b) rotation of line elements along the centerline  $\alpha(x,t)$.}
\label{fig3}
\end{figure}

In this section for showing the accuracy of the controller designed by the boundary control method, replacing real values instead of parameters and using finite element modeling that was presented in the previous section, the strain gradient Timoshenko micro-cantilever for two cases before and after applying the control actuator is simulated. 

First of all, it is required that the system parameters become nondimensionalized. The following nondimensionalized variables and parameters are utilized.
\begin{align}
    \left\{ \begin{matrix}
   \begin{matrix}
   \tilde{x}=\frac{x}{L}  \\
   \begin{matrix}
   \tilde{L}=\frac{L}{L}  \\
   \begin{matrix}
   \tilde{b}=\frac{b}{L}  \\
   \begin{matrix}
   {{{\tilde{h}}}^{b}}=\frac{{{h}^{b}}}{L}  \\
   {{{\tilde{h}}}^{p}}=\frac{{{h}^{p}}}{L}  \\
\end{matrix}  \\
\end{matrix}  \\
\end{matrix}  \\
\end{matrix} & \begin{matrix}
   \begin{matrix}
   {{{\tilde{\rho }}}^{b}}=\frac{{{\rho }^{b}}}{{{\rho }^{b}}}  \\
   \begin{matrix}
   {{{\tilde{\rho }}}^{p}}=\frac{{{\rho }^{p}}}{{{\rho }^{b}}}  \\
   \begin{matrix}
   {{{\tilde{I}}}^{b}}=\frac{{{I}^{b}}}{{{L}^{4}}}  \\
   \begin{matrix}
   {{{\tilde{I}}}^{p}}=\frac{{{I}^{p}}}{{{L}^{4}}}  \\
   {{{\tilde{c}}}_{ij}}=\frac{{{c}_{ij}}}{{{\rho }^{b}}{{L}^{2}}\omega _{1}^{2}}  \\
\end{matrix}  \\
\end{matrix}  \\
\end{matrix}  \\
\end{matrix} & \begin{matrix}
   \begin{matrix}
   {{{\tilde{e}}}_{13}}=\frac{{{e}_{13}}}{{{e}_{13}}}  \\
   \begin{matrix}
   {{{\tilde{c}}}_{1}}=\frac{{{c}_{1}}}{{{\rho }^{b}}{{L}^{2}}{{\omega }_{1}}}  \\
   \begin{matrix}
   {{{\tilde{c}}}_{2}}=\frac{{{c}_{2}}}{{{\rho }^{b}}{{L}^{2}}{{\omega }_{1}}}  \\
   \begin{matrix}
   \tilde{u}=\frac{u{{e}_{13}}}{{{\rho }^{b}}{{L}^{3}}\omega _{1}^{2}}  \\
   \tilde{t}=t{{\omega }_{1}}  \\
\end{matrix}  \\
\end{matrix}  \\
\end{matrix}  \\
\end{matrix} & \begin{matrix}
   {{{\tilde{l}}}_{0}}=\frac{{{l}_{0}}}{L}  \\
   \begin{matrix}
   {{{\tilde{l}}}_{1}}=\frac{{{l}_{1}}}{L}  \\
   \begin{matrix}
   {{{\tilde{l}}}_{2}}=\frac{{{l}_{2}}}{L}  \\
   \begin{matrix}
   {{{\tilde{E}}}^{b}}=\frac{{{E}^{b}}}{{{E}^{b}}}  \\
   {{{\tilde{E}}}^{p}}=\frac{{{E}^{p}}}{{{E}^{b}}}  \\
\end{matrix}  \\
\end{matrix}  \\
\end{matrix}  \\
\end{matrix}  \\
\end{matrix}  \\
\end{matrix}  \\
\end{matrix} \right.
\label{eq65}
\end{align}

Also, the physical characteristics of the beam and piezoelectric layer can be found in \textcolor{blue}{Table.\ref{table1}}\textcolor{blue}{\cite{shirazi2013tip, maluf2002introduction, gad2005mems}} and the geometry of the piezoelectric and beam are given in \textcolor{blue}{Table.\ref{table2}}\textcolor{blue}{\cite{shirazi2013tip}}.

According to table 1, we have:
\begin{align}
    {{e}_{13}}=\underset{1}{\overset{3}{\mathop \sum }}\,{{d}_{3i}}{{c}_{1i}}=-3.621{}^{C}/{}_{{{m}^{2}}}
    \label{eq66}
\end{align}

Also, the bulk module, shear module and the shear coefficient of the Timoshenko beam are obtained from equations \eqref{eq67}, \eqref{eq68} and \eqref{eq69}.
\begin{align}
    K=\frac{E}{3\left( 1-2\upsilon  \right)}
    \label{eq67}\\
    \mu =\frac{E}{2\left( 1+\upsilon  \right)}
    \label{eq68}\\
    {{K}_{s}}=\frac{5+5\upsilon }{6+5\upsilon }
    \label{eq69}
\end{align}

\begin{figure}
\centering
\includegraphics[width=80mm,scale=0.5]{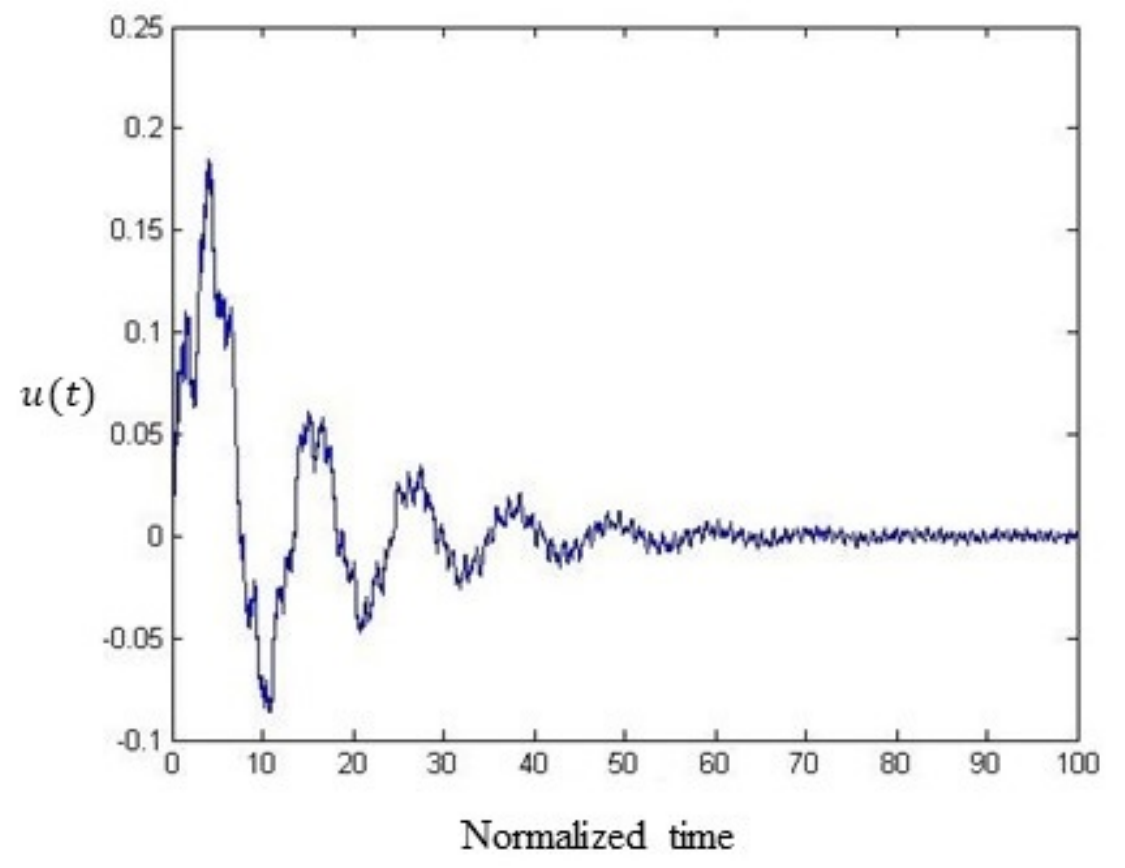}
\caption{Control voltage  $u(t)=k_u \alpha_t(L)$.}
\label{fig4}
\end{figure}

In the equation \eqref{eq28}, a proper control gain obtained via trial and error which has a suitable settling time and transient response is selected as $k_u=0.6$ . By replacing this value in equations and considering ten nodes on the beam, the equation \eqref{eq64} is solved for two mentioned states in a distinct time period. The results are shown in \textcolor{blue}{Fig.\ref{fig2}} and \textcolor{blue}{Fig.\ref{fig3}}, and the controller voltage that was obtained in the equation \eqref{eq28} is shown in \textcolor{blue}{Fig.\ref{fig4}}. 
\begin{table}[]
\caption{Material properties of silicon dioxide beam and PZT actuator.}
\label{table1}
\centering
\begin{tabular}{lll}
\hline\hline
Material & SiO2 & PZT \\ \hline \hline
Density (Kg/m3) & 2200 & 7700 \\
Poisson coefficient & 0.17 & 0.31 \\
Young modulus of   elasticity (GPa) & 73 & 71 \\
Piezoelectric      Constants (10-12C/N) & - & \begin{tabular}[c]{@{}l@{}}d31=175\\ d33=400\\ \\ d55=580\end{tabular} \\
Relative permittivity & 3.9 & 1700 \\ \hline
\end{tabular}%
\end{table}
\begin{table}[]
\centering
\caption{Geometrical dimensions of beam and piezoelectric layer (all in µm)}
\label{table2}
\begin{tabular}{ll}
\hline\hline
Beam length & 90 \\ \hline\hline
Beam thickness & 10 \\
Beam width & 30 \\
Piezoelectric length & 150 \\
Piezoelectric thickness & 10 \\
Piezoelectric width & 30 \\ \hline
\end{tabular}%
\end{table}
\section{Simulation}
As it is clear, after applying the control action, the vibration of the system caused by the non-zero initial displacement has been suppressed and the system has become asymptotically stable.Such vibration damping mechanism can be also realized for the acoustic wave utilizing the
destructive interferences\textcolor{blue}{\cite{ghaffarivardavagh2019ultra}}. According to this simulation, the accuracy of control law and obtained equations can be confirmed. This method of control is used for different applications such as vibration control of the fluid containers\textcolor{blue}{\cite{mehrvarz2019vibration}}. If strain gradient model is used for flexible structure in side-wall of these containers, better results will be obtained.  

\section{Conclusion}
In this paper strain gradient Timoshenko micro-cantilever with a piezoelectric layer laminated on one side of the beam was modeled and equations of the system with boundary conditions were obtained in state space. Then well-posedness of equations were checked and by using the Lyapunov function and LaSalle’s invariant set theorem, a control law for the stability of the system was proven. This control law for suppressing the vibration of the system was achieved from the feedback of temporal derivatives of boundary states of the beam, and it was applied through exciting voltage of the piezoelectric layer. For showing accuracy of the designed controller, the simulation was done. In this work by using finite element method and strain gradient Timoshenko element, the system was modeled and by using numerical solution for two cases means closed-loop and open loop systems, the simulation was performed which verified the achieved theoretical results of this work.
\bibliographystyle{IEEEtran}
\bibliography{main.bib}

\end{document}